\documentclass[conference]{IEEEtran}
\pdfoutput=1
\usepackage{cite}
\usepackage[utf8]{inputenc}
\usepackage{xcolor}
\usepackage{amsmath,amssymb,amsfonts}
\usepackage{graphicx}
\usepackage{textcomp}
\usepackage{tabularx}
\usepackage{booktabs} 
\usepackage{xcolor}
\usepackage{enumerate}
\def\BibTeX{{\rm B\kern-.05em{\sc i\kern-.025em b}\kern-.08em
    T\kern-.1667em\lower.7ex\hbox{E}\kern-.125emX}}
    
\usepackage{placeins}

\usepackage{bbm}
\usepackage{verbatim}
\usepackage{amsthm}
\usepackage{bm}
\usepackage{subcaption}
\usepackage[linesnumbered]{algorithm2e}
\newcommand{\bb}[1]{\mathbb{#1}}
\let\bf\relax
\newcommand{\bf}[1]{\mathbf{#1}}
\let\rm\relax
\newcommand{\rm}[1]{\mathrm{#1}}
\let\cal\relax
\newcommand{\cal}[1]{\mathcal{#1}}

\usepackage{ulem}
\usepackage[xcolor,pdftex]{changebar}

\DeclareMathOperator*{\argmax}{arg\,max}

\theoremstyle{definition}

\usepackage{comment} 
\usepackage{tikz}
\usetikzlibrary{calc}
\usetikzlibrary{matrix}
\usepackage{todonotes}

\title{Joint Slot and Power Optimization for Grant Free Random Access with Unknown and Heterogeneous Device Activity}

\author{\IEEEauthorblockN{Alix Jeannerot}
\IEEEauthorblockA{\textit{INSA Lyon, Inria, CITI, UR3720} \\
69621 Villeurbanne, France \\
alix.jeannerot@inria.fr}
\and
\IEEEauthorblockN{Malcolm Egan}
\IEEEauthorblockA{\textit{Inria, INSA Lyon, CITI, UR3720} \\
69621 Villeurbanne, France \\
malcolm.egan@inria.fr}
\and
\IEEEauthorblockN{Jean-Marie Gorce}
\IEEEauthorblockA{\textit{INSA Lyon, Inria, CITI, UR3720} \\
69621 Villeurbanne, France \\
jean-marie.gorce@insa-lyon.fr}
}

\begin{document}

\maketitle

\begin{abstract}
Grant Free Random Access (GFRA) is a popular protocol in the Internet of Things (IoT) to reduce the control signaling. GFRA is a framed protocol where each frame is split into two parts: device identification; and data transmission part which can be viewed as a form of Frame Slotted ALOHA (FSA). A common assumption in FSA is device homogeneity; that is the probability that a device seeks to transmit data in a particular frame is common for all devices and independent of the other devices. Recent work has investigated the possibility of tuning the FSA protocol to the statistics of the network by changing the probability for a particular device to access a particular slot. However, power control with a successive interference cancellation (SIC) receiver has not yet been considered to further increase the performance of the tuned FSA protocols. In this paper, we propose algorithms to jointly optimize both the slot selection and the transmit power of the devices to minimize the outage of the devices in the network. We show via a simulation study that our algorithms can outperform baselines (including slotted ALOHA) in terms of expected number of devices transmitting without outage and in term of transmit power.
\end{abstract}


\section{Introduction}
The selection of transmit resources with limited coordination between devices and with the base station is a key problem in wireless multiple access networks which lead to a family of protocols known as ALOHA. Slotted ALOHA \cite{robertsALOHAPacketSystem1975} protocols and most of its variants such as framed slotted ALOHA (FSA) \cite{wieselthierExactAnalysisPerformance1989b}, coded slotted ALOHA (CSA) \cite{paoliniCodedSlottedALOHA2015}, irregular repetition slotted ALOHA (IRSA) \cite{livaGraphBasedAnalysisOptimization2011}, assume that the network is homogeneous, meaning that devices are independent and equally likely to transmit data. However within the context of modern communication networks like Internet of Things or event-driven communication, the probability that a device is active in a specific frame is likely to vary from device to device.

A key question is whether exploiting knowledge of device heterogeneity can improve resource utilization. To this end, new variants of Frame Slotted ALOHA have been proposed, where the probability that a particular device chooses a particular slot is not uniform but is optimized as a function of the statistics of the network. In \cite{kalorRandomAccessSchemes2018a, raghuwanshiChannelSchedulingIoT2024}, stochastic resource selection is optimized based on the joint probability of activity of the devices. Optimization is based on heuristics in \cite{kalorRandomAccessSchemes2018a} and by solving a quadratic program in \cite{raghuwanshiChannelSchedulingIoT2024}. However both approaches require the knowledge at the base station of the activity probability of the devices in the network. To relax this assumption, the work in \cite{zhengStochasticResourceOptimization2021,zhengStochasticResourceAllocation2022} proposed to sequentially optimize the stochastic resource allocation based only on the devices active in each frame, achieved via projected stochastic gradient descent. Finally, in \cite{jeannerotMitigatingUserIdentification2023, jeannerotExploitingDeviceHeterogeneity2024}, the impact of imperfect estimation of device activity (e.g., via device identification algorithms \cite{chetotJointIdentificationChannel2021}) on the optimization is studied as well as method to mitigate the effect of activity estimation errors. 

Collisions arising from multiple devices utilizing the same resources in Slotted ALOHA is a limiting factor when seeking to increase the density of the network. To this end, it is common to assume that the base station is able to exploit successive interference cancellation (SIC). In SIC, the data of each device is sequentially decoded, and subtracted from the received signal in order to increase the SINR of the devices that are yet to be decoded. SIC was used in \cite{zhengStochasticResourceAllocation2022, zhengStochasticResourceOptimization2021} to take advantage of the different channel conditions to further reduce the contention in network. However the power was assumed to be the same for all the devices. To further exploit the SIC receiver, it is desirable to also tune the power at which the devices transmit in addition to the slot they transmit in. As such, power control can be exploited in a similar manner as power domain non orthogonal multiple access (PD-NOMA) \cite{mounirSelectionPowerAllocation2022}.

In this paper, we propose an algorithm in the context of NB-IoT networks \cite{hoglundOverview3GPPRelease2017} to jointly optimize two different types of resources: the temporal slot (orthogonal) and the transmit power (non-orthogonal).
Our contributions are as follows: 
\begin{enumerate}[(i)]
    \item we propose an algorithm to jointly optimize the slot selection and the transmit power of devices to minimize the outage probability of the devices, with a SIC receiver at the base station. In contrast, existing work \cite{zhengStochasticResourceOptimization2021,zhengStochasticResourceAllocation2022} exploiting SIC assumes constant power levels.
    \item We propose an additional algorithm to adapt power control resulting from the algorithm in (i), leading to further reductions of the power consumption without reducing throughput performance. This algorithm is run after the algorithm in (i) has completed, allowing to minimize the transmit power when the network runs. 
    \item We show via a simulation study that our algorithms can exploit heterogeneous activity, leading either to improvements in throughput performance or reductions in energy consumption. 
\end{enumerate}

In section \ref{sec:model} we present our system model, in section \ref{sec:obj} we define the objective function, in section \ref{sec:opti} we introduce the optimization method and finally we show simulation results in section \ref{sec:simul}. 

\section{System model}
\label{sec:model}
Consider a network consisting of a single base station with $N_r$ antennas and $N$ devices. In NB-IoT, uplink communications occur over a single carrier that is shared by the devices. Devices are assumed to be synchronized and the time is split into frames,  each consisting of $K$ slots. Each frames consists of two parts: device identification and data transmission. During the device identification part, all devices that seek to transmit data send their unique identifier (pilot) at same time. Data transmission consists of $K$ slots, active devices can transmit data on only one slot per frame. Each slot is further divided into $L$ symbols.

\subsection{Activity model}
In each frame only a subset of the $N$ devices is active. The probability that a particular device $i$ is active in a frame is $p_i$ and let $X_i\in\{0,1\}$, the state of the device (respectively inactive or active), $X_i\sim\rm{Ber(p_i)}$. We define $\bf{X}\in\{0,1\}^N$ the vector of active devices and $\bf{p}=[p_1\ldots p_N]$ the vector of activity probabilities, $\bf{X}\sim p_{\bf{X}}$, where $p_{\bf{X}}$ is $\rm{Ber}(\bf{p})$; without loss of generality, we assume devices are sorted in terms of activity: $p_1\leq\ldots\leq p_N$. As such, devices are mutually independent with different marginals:
\begin{align}
p_{\bf{X}}(\bf{x})=\prod_{i=1}^Np_i^{x_i}(1-p_i)^{1-x_i},\label{eq:bernouilli}.
\end{align}
The activity vector $\bf{X}^t$ in frame $t$ is independent of the other frames;  \textit{i.e.}, $\mathbf{X}^t$ is independent of $\mathbf{X}^{t^\prime}$ $\forall t\neq t^\prime$.
We consider the following assumptions:
\begin{enumerate}[(i)]
    \item the base station does not have knowledge of the parameter $\bf{p}$ of the activity distribution,
    \item we assume that, during the pilot phase, the base station detects without error the active devices (\textit{e.g} using receiver like \cite{chetotJointIdentificationChannel2021}) and thus has access to sample $\bf{X}^t$.
\end{enumerate}

\subsection{Slot selection}
In contrast to standard approaches that utilize FSA to perform the slot selection, we follow the line of work \cite{kalorRandomAccessSchemes2018a, raghuwanshiChannelSchedulingIoT2024,zhengStochasticResourceAllocation2022, zhengStochasticResourceOptimization2021, jeannerotMitigatingUserIdentification2023,jeannerotExploitingDeviceHeterogeneity2024} that perform the slot selection according to a slot allocation matrix $\bf{A}\in \bb{R}_+^{N \times K}$. Each element of $\bf{A}$ represents the probability that a device selects a particular slot when it is active. More precisely: $a_{i,k}=\rm{Pr}\text{(device $i$ selects slot $k$ $|$ $X_i=1$)}$. In other words, matrix $\bf{A}$ must satisfy the constraint $\mathcal{H} = \{\bf{A} \in \bb{R}^{N \times K}_+: \sum_{k=1}^K A_{ik} = 1,~i=1,\ldots,N\}$. Note that in the standard FSA protocol, the allocation matrix is the constant matrix $\bf{A}_{\rm{ALOHA, ij}}=\frac{1}{K}$

\subsection{Transmit power}
The transmit power of the devices is $\bf{P}=[P_{1},\ldots,P_{N}]\in\mathbb{R}_+^{N}$. The transmit power of each individual device is constrained by $P_{\min,i}\leq P_i\leq P_{\max}$, where $0<P_{\min,i}\leq P_{\max}$, $P_{\min,i}$ should be defined to be large enough to ensure reliable reception if no devices, other than device $i$, are transmitting, and $P_{\max}$ is a value that depends on the standard or other regulations. The choice of the transmit power is independent of the slot.

\subsection{Channel model and receiver}
We assume that devices are equipped with a single transmit antenna and that the base station is equipped with $N_r$ receive antennas. The channel between the devices and the base station is modeled by block Rayleigh fading with additive white Gaussian noise. During a slot, devices send a fixed number of symbols $L$.
In the slot $s$ of frame $t$, assuming there are $N_{a,s}$ active devices, the received signal $\bf{Y}^t_s\in\bb{C}^{L\times N_r}$ is:  
\begin{align}
    \bf{Y}^t_s&=\sqrt{\rm{diag}(\bf{P})}\bf{D}^t(\bf{H}^t)^\dag + \bf{W}^t,
    \label{eq:channel}
\end{align}
where $\rm{diag}(\bf{P})$ represents the $N_{a,s}\times N_{a,s}$ diagonal matrix of device's transmit power, $\bf{D}^t=[(\bf{d}_i^t)^\dagger,\ldots,(\bf{d}_{N_{a,s}}^t)^\dagger]\in\bb{C}^{L\times N_{a,s}}$ with $\bb{E}[\|\bf{d}_i\|^2]=1$ and represents the data symbols sent by the active devices, $\mathbf{H}^t=[(\bf{h}_i^t)^\dag,\ldots,(\bf{h}_{N_{a,s}}^t)^\dag]\in\bb{C}^{N_r\times N_{a,s}}$ is the channel gain matrix, assuming the channel is constant throughout over frame $t$, and $\bf{W}^t\in\bb{C}^{L\times N_r}$ is circularly Gaussian noise $\sim\mathcal{CN}(0,I\sigma^2)$.

In this work, perfect channel state information at the receiver (CSIR) is assumed and a MRC spatial filter is utilized to maximize the SINR of the devices. The signal associated to device $i$ is (superscript $t$ is dropped to ease the notation):
\begin{align*}
    \hat{\bf{y}}_i=\bf{Y}\frac{\bf{h}_i}{\|\bf{h}_i\|^2}=\sqrt{P_i}\bf{d}_i+\sum_{j\neq i}\sqrt{P_j}\bf{d}_j\frac{\bf{h}_j^\dagger\bf{h}_i}{\|\bf{h}_i\|^2}+\hat{\bf{w}}_i.
\end{align*}
Where $\hat{\bf{y}}_i\in\bb{C}^L$ is the received the approximation of $\bf{d}_i$, and $\hat{\bf{w}}\sim\cal{CN}\left(0,\rm{diag}\left(\frac{1}{\|\bf{h_i}\|^2}\right)\sigma^2\right)$.

For a subset of devices $\mathcal{S}\subset\{1,\ldots N\}$ ordered in decreasing received power, and a given power vector $\bf{P}$, we can express the SINR of the first device $S(1)$ in the set $\mathcal{S}$ under the noise and the interference of the other devices of $\mathcal{S}$: 
\begin{align}
    \rm{SINR}(\mathcal{S},\bf{P})=\frac{P_{\mathcal{S}(1)}}{\frac{\sigma^2}{\|\bf{h}_{\mathcal{S}(1)}\|^2}+\sum_ {i=2}^{|\mathcal{S}|}\left(\frac{|\bf{h}_{\mathcal{S}(i)}^\dagger\bf{h}_{\mathcal{S}(1)}|}{\|\bf{h}_{\mathcal{S}(1)}\|^2}\right)^2P_{\mathcal{S}(i)}}.\label{eq:sinr}
\end{align}
We assume that the data of $\mathcal{S}(1)$ can be decoded if $\rm{SINR}(\mathcal{S},\bf{P})>\gamma;~\gamma>0$. An outage arises when $\rm{SINR}(\mathcal{S},\bf{P})\leq\gamma$. We further assume that the receiver performs successive interference cancellation (SIC): if $\rm{SINR}(\mathcal{S},\bf{P})>\gamma$ the receiver can proceed to compute $\rm{SINR}(\mathcal{S}\setminus{\mathcal{S}(1)},\bf{P})$, otherwise all devices in $\mathcal{S}$ are considered outed of the network.

\section{Objective}
\label{sec:obj}

In this paper, we are interested in maximizing the number of devices that transmit without an outage \cite{zhengStochasticResourceAllocation2022}. More precisely we aim to maximize the expected number of devices for which the MRC receiver combined with the SIC decoder will lead to a SINR that is sufficient to be able to recover its sent signal. This objective is defined as an expectation over the activity distribution of the device $p_{\bf{X}}$ defined in (\ref{eq:bernouilli}):

\begin{align}
    T(\bf{A,P})&=\bb{E}_{\bf{X}\sim p_{\bf{X}}}[T(\bf{A,P;X})]\label{eq:obj}\\
    T(\bf{A,P;X})&=\sum_{k = 1}^K {\sum\limits_{\mathcal{S} \in \mathcal{P}(N)} {{Q_k}}} (\bf{A},\mathcal{S}\mid {\mathbf{X}})\rm{SIC}(\mathcal{S}, \bf{P}) \label{eq:obj_x},
 \end{align}
 where $\mathcal{P}(N)$ represents the power set of $\{1,\ldots, N\}$,
\begin{align}
    {Q_k}(\bf{A},\mathcal{S}\mid {\mathbf{X}})&=\prod\limits_{i \in \mathcal{S}} {{X_i}} {A_{ik}}\prod\limits_{j \in {\mathcal{S}^c}} {\left( {1 - {X_j}{A_{jk}}} \right)},\label{eq:q}
\end{align}
represent the probability that all devices in a subset of the devices are choosing a particular slot while the other devices chose another slot, and:
\begin{align}
    &\rm{SIC}(\mathcal{S},\bf{P})=\sum_{l = 1}^{|\mathcal{S}|-1}\mathbbm{1}\left\{\rm{SINR}(\mathcal{S},\bf{P})  > \gamma \right\}\nonumber\\ 
    &~~~~~\cdot\prod_{m=1}^l\mathbbm{1} \left\{ \rm{SINR}(\mathcal{S}\setminus\{\mathcal{S}(1),\ldots,\mathcal{S}(m)\},\bf{P})  > \gamma \right\}\label{eq:sic},
\end{align}
represents the number of devices within a set of device $\mathcal{S}$ that have a sufficient SINR to be decoded under a SIC decoder. We assume that, all sets $\mathcal{S}\in\mathcal{P}(N)$, are sorted in terms of decreasing received power. Note that if a device $\mathcal{S}(i)$ cannot be decoded then all subsequent devices $\mathcal{S}(j),~i<j$, also cannot be decoded.

To allow for a fair comparison of the performance under different network statistics, we define the normalized expected number of devices that transmit without outage:
\begin{align}
    T^N(\bf{A,P})&=\frac{\bb{E}_{\bf{X}\sim p_{\bf{X}}}[T_1(\bf{A,P;X})]}{\sum_{i=1}^Np_i}\label{eq:obj_norm}.
\end{align}
 
\section{Proposed Algorithms}
\label{sec:opti}
\subsection{Slot and power optimization}
Optimization of $\bf{A}$ and $\bf{P}$ in (\ref{eq:obj}) is performed in an sequential manner using stochastic optimization based on the vector of active devices $\bf{X}^t$ of each frame $t$. The optimization problem is written as follows:
\begin{align}
    \bf{A}^\star,\bf{P}^\star&=\argmax_{\substack{\bf{A}\in\bb{R}^{N\times K}_+,\\\bf{P}\in\bb{R}^N}} T_1(\bf{A,P})\label{eq:opti_pb}\\
    &\text{s.t.:~} P_{\rm{min},i}\leq P_i\leq P_\rm{max},~\forall i\in\{1\ldots N\}\nonumber\\
    &\sum_j A_{i,j=1},~\forall i\in\{1\ldots N\}\nonumber.
\end{align}
Note that problem (\ref{eq:opti_pb}) is non convex, due to the nature of (\ref{eq:q}). Note also that the objective in (\ref{eq:opti_pb}) is non-differentiable, due to the indicator function in (\ref{eq:sic}), to circumvent the latter issue, we propose to approximate the indicator with a properly scaled sigmoid function centered on $\gamma$:
\begin{align}
    &\widetilde{\rm{SIC}}(\mathcal{S},\bf{P})=\sum_{l = 1}^{|\mathcal{S}|-1}\sigma_f\left(b\rm{SINR}(\mathcal{S},\bf{P})  - \gamma \right)\nonumber\\ 
    &~~~~\cdot\prod_{m=1}^l\sigma_f\left(b\rm{SINR}\left(\mathcal{S}\setminus\{\mathcal{S}(1),\ldots,\mathcal{S}(m)\},\bf{P}\right)  - \gamma \right)\label{eq:sic_tilde},
\end{align}
where $\sigma_f(x)=\frac{1}{1-e^{-x}}$ is the sigmoid function and b is a sharpness parameter. Let:
\begin{align}
    \tilde{T}(\bf{A,P;X})&=\sum_{k = 1}^K {\sum\limits_{\mathcal{S} \in \mathcal{P}(N)} {{Q_k}}} (\bf{A},\mathcal{S}\mid {\mathbf{X}})\widetilde{\rm{SIC}}(\mathcal{S}, \bf{P}) \label{eq:obj_x_tilde}.
\end{align}
In each frame, by detecting active devices, we obtain a sample $\bf{X}^t\sim p_{\bf{X}}$ yielding $\tilde{T}(\bf{A}^t,\bf{P}^t;\bf{X}^t)$. 
Considering that the two optimization variables $\bf{A}$ and $\bf{P}$ have different impact and that the activity of the devices is heterogeneous, we use \texttt{ADAGRAD} \cite{duchiAdaptiveSubgradientMethods2011}. The optimization algorithm is detailed in Alg. \ref{alg:adagrad}, where \texttt{ADAGRAD} is as implemented in the Pytorch package \cite{Paszke_PyTorch_An_Imperative_2019} and where the operator $\Pi_[\mathcal{H}]\{\cdot\}$ represent the projection on the constraint set of the parameters.


\RestyleAlgo{ruled}
\begin{algorithm}
    \caption{Optimization of $\bf{A}$ and $\bf{P}$ to solve (\ref{eq:opti_pb}).}
	\label{alg:adagrad}
    Choose initial allocation matrix $\mathbf{A}^1$ initial power allocation $\bf{P}^1$, and step-size sequence $\{\mu^t\}$\\
	$t \leftarrow 1$.\\
	\While {not converged}{
        Detect active devices to obtain $\bf{X}^t$\\
        $(\mathbf{A}^{t+1},\bf{P}^{t+1})\leftarrow $\texttt{ADAGRAD}$(\mu^t,(\bf{A}^t,\bf{P}^t),\tilde{T}(\bf{A}^t,\bf{P}^t;\bf{X}^t))$\\
        $\bf{A}^{t+1}\leftarrow\Pi_\mathcal{H}[\mathbf{A}^{t+1}]$\\
        $\bf{P}^{t+1}\leftarrow\Pi_{[P_{\min,i}, P_{\max}]}[\mathbf{P}^{t+1}]$\\
    }
\end{algorithm}
 
\subsection{Power Reduction}
For a fixed allocation matrix $\bf{A}$ and a given objective value, there are several power vectors that achieve this value. Hence, after (\ref{eq:opti_pb}) is solved via Alg.~\ref{alg:adagrad}, we seek to minimize the transmit power of all the devices without decreasing the objective. A way to achieve this is, for all sets $\mathcal{S}$ for which $\rm{SINR}(\mathcal{S},\bf{P})>\gamma$,to find the smallest transmit power vector that will lead to a SINR that is just above the threshold $\gamma$. Then for all devices the maximum value over all the minimal transmit power is taken. This process, described in Alg.~\ref{alg:cleanup}, is repeated until the resulting power vector cannot be reduced further. 

\RestyleAlgo{ruled}
\begin{algorithm}[ht!]
\caption{Reducing the power}                 
\label{alg:cleanup}
\SetKwProg{Fn}{Function}{}{}
\SetKwFunction{power}{power\_cleanup}

\Fn{\power{$\mathcal{S},\bf{P}, \bf{h}$}}{
    $sinr, I=SINR(\mathcal{S}, \bf{P}, \bf{h})$\tcp*{I is the denominator of (\ref{eq:sinr})}
    \If{$sinr > \gamma$}{
        \tcp{Find min power for $\mathcal{S}$(1)}
        $p_{\rm{min}}=(\gamma*I)/(|h[1]|)^2+\epsilon$\tcp*{$\epsilon>0$}

	\If{order of SIC decoding is changed}{
	    $p_{min}=$ min value that do not change the order
	}
	return $p_{min}$
    }
    \Else{
        \tcp{Iterate the other combination of devices and take the max}
            $\mathbf{P}_\rm{temp}=[0,\ldots,0]$

            \For {$i=2;~i\leq|\mathcal{S}|;~i++$}{
                $\mathcal{S}_{\rm{next}}=\mathcal{S}\setminus\{\mathcal{S}(i)\}$\\
                $\bf{P}_\rm{temp}[i]=\power(\mathcal{S}_\rm{next},\bf{P}_{\rm{temp}},\bf{h})$
            }
            return $\max(\bf{P}_\rm{temp})$
    }
}
$\bf{P}_{\rm{cleaned}}=[0,\ldots,0]$

\For {$k=0;~k<K;~k++$}{
    $\mathcal{S}=\text{indices where}(A[:,k]>0)$\tcp*{Devices transmitting in k}
    $\mathcal{S}$=\text{sort}($\mathcal{S}$, $\bf{P}$)\tcp*{By decreasing power}

    $P_k=\power(\bf{P}, g, \mathcal{S})$\\
    $P_{\rm{cleaned}}[\mathcal{S}(1)]=\max(P_{\rm{cleaned}}[\mathcal{S}(1)], P_k)$

    \For {$i=2;~i\leq|\mathcal{S}|;~i++$}{
        $\mathcal{S}\leftarrow\mathcal{S}\setminus\{\mathcal{S}(i-1)\}$\\
            $P_k=\power(\mathcal{S}, \bf{P}, g)$\\
    $P_{\rm{cleaned}}[\mathcal{S}(1)]=\max(P_{\rm{cleaned}}[\mathcal{S}(1)], P_k)$
    }
}
return $P_{\mathrm{cleaned}}$
\end{algorithm}

\section{Numerical results}
\label{sec:simul}
\subsection{Methods and Baselines}
\subsubsection{Independent activity}
We evaluate the performance of a slot and a power allocation optimized with Alg.\ref{alg:adagrad}. Three methods are considered: 
\begin{enumerate}[(i)]
    \item Optimization with Alg. \ref{alg:adagrad} only.
    \item Optimization with Alg. \ref{alg:adagrad} followed by power reduction algorithm Alg. \ref{alg:cleanup}.
\item Optimization with Alg. \ref{alg:adagrad} with $\ell^1$ regularization: the objective optimized is $\widetilde{T}(\bf{A}^t\bf{P}^t)+\lambda\|\bf{P^t}\|_1$, with $\lambda=0.001$.
\end{enumerate}

\subsubsection{ALOHA with structured power levels}
\label{sec:aloha_struc}
The devices are choosing their transmit slot uniformly at random (according to the ALOHA matrix) $\bf{A}_{\rm{ALOHA},~ij}=\frac{1}{K},~\forall{i,j}$. To cover a sufficient power diversity, a worst case scenario for the definition of the power levels is as follows:
\begin{enumerate}[(i)]
    \item set the first level $P_{l1}=\max_iP_{\min,i}$ to ensure that the device with the worst channel condition can transmit,
     \item define power levels of order $j$ as $P_{lj}=2^jP_{l1}$, $\forall~j$ such that $P_{lj}\leq P_{\max}$.
\end{enumerate}
Devices choose their transmit slots according to $\bf{A}_{\rm{ALOHA}}$ and the transmit power vector is $\bf{P}=[P_{l1},\ldots,P_{lj},P_{l1}\ldots]$ (power levels are assigned cyclically among devices).

\subsubsection{Greedy allocation}
When the activity probability is known at the BS, a greedy allocation can be considered. It ensures that devices the most active can be decoded by assigning them either orthogonal slots or different transmit power levels at the expense of the least active devices that share a single slot and a single power level. The greedy allocation places the least probable devices in the same slot at transmit power $P_{lj}$, and then by orthogonally grouping blocks of $K$ devices on $K$ slots at the lower power levels. Matrix $\bf{A_{\rm{greedy}}}$ can be represented using blocks of diagonal matrices $I_k$ of size $K$. The allocation gives:
  \begin{equation*}
  \begin{split}
    \bf{A}_{\rm{greedy}}=\begin{bmatrix}1&0&\dots&0\\&\vdots&\\1&0&\dots&0\\&I_k&\\&\vdots&\\&I_k&\end{bmatrix}
  \end{split}
  ,~
  \begin{split}
      \bf{P}_{\rm{greedy}}=\begin{bmatrix}P_{lj}\\\vdots\\P_{lj}\\P_{lj}\\\vdots\\ P_{l1}\end{bmatrix}
  \end{split}
\end{equation*} 
\subsection{Simulation setup}
We consider a network consisting of $N$ devices and $K$ slots. In each frame, devices are either active or inactive, independently of the other devices and of the frame number, according to a Bernouilli law of parameter $\bf{p}=[p_1 \ldots p_N]$. In each frame $t$ the base station has only access to a sample $\bf{X}^t\sim\rm{Ber}(\bf{p})$. The channels are assumed to be fixed during all the simulation, $\bf{H}^t=\bf{H},~\forall t$. We consider a Rayleigh fading model where the channel between device $i$ and antenna $l$ of the base station is $h_{i,l}\sim\mathcal{CN}(0,1)$. The base station is equipped with $N_r=2$ receive antennas.

We consider 4 different scenarios covering different number of slots, devices and activity probabilities. 
\begin{enumerate}
    \item $N=5$, $K=2$, $\bf{p}=[.08, .12, .33, .35, .38]$
    \item $N=5$, $K=2$, $\bf{p}=[.22, .38, .41, .70, .88]$
    \item $N=8$, $K=3$, $\bf{p}=[.10, .12, .37, .41, .41, .41, .42, .45]$
    \item $N=8$, $K=3$, $\bf{p}=[.36, .38, .39, .46, .54, .64, .68, .83]$
\end{enumerate}
Alg. \ref{alg:adagrad} is run for 100000 frames with $\mu^t=0.01~\forall t$, the initial allocation matrix $\bf{A}^1$ is random and the initial power allocation is $\bf{P}^1=[P_{\min,1},\ldots, P_{\min,N}]$, $P_{\max}$ is 6, and the number of power level used by the baselines is 2. The sharpness parameter in (\ref{eq:sic_tilde}) is $b=10$.

\subsection{Results}

\begin{figure}
    \centering
    \includegraphics[width=\linewidth]{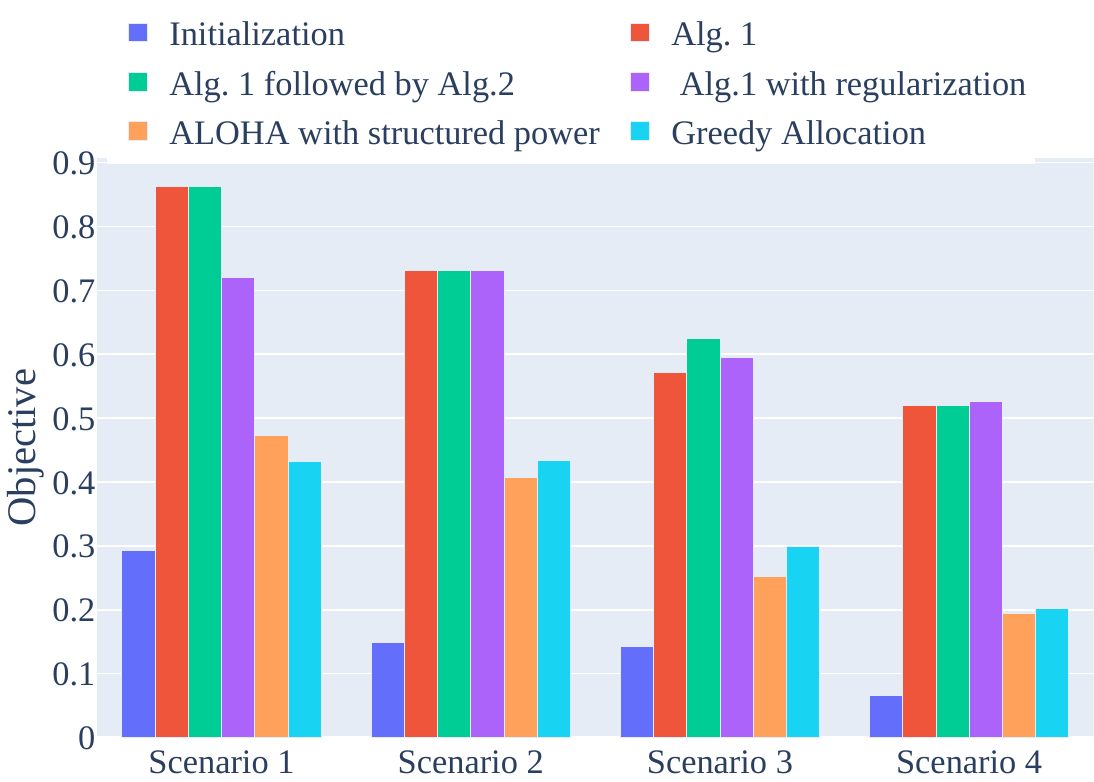}
    \caption{Resulting objective $T^N(\bf{A,P})$ for the different method in the each scenarios.}
    \label{fig:outage}
\end{figure}
\begin{figure}
    \centering
    \includegraphics[width=\linewidth]{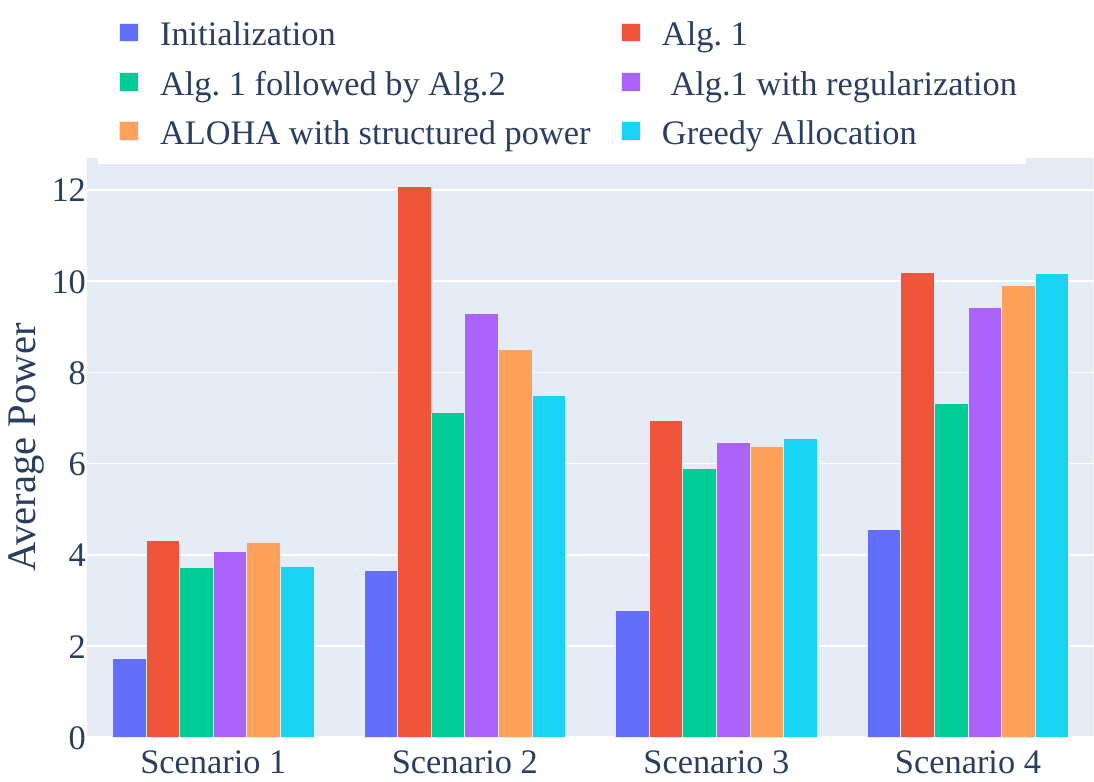}
    \caption{Average power ($\bb{E}_{\bf{X}}[\bf{P}^T\bf{X}]$) used.}
    \label{fig:power}
\end{figure}

Our simulations compare the resulting expected number of devices able to transmit without outage and the expected transmit power of devices in the network. The first bar in dark blue represent the performance obtained by the initial $\bf{A}^1$ and $\bf{X}^1$. In Fig.\ref{fig:outage}, our methods (red, green and purple, resp. 2nd, 3rd and 4th bar) are able to find a power and a slot allocation that allows to increase by up to 150\% the objective $T^N(\bf{A},\bf{P})$ (in scenario 4) compared to the baseline considered (orange and light blue, resp. 5th and 6th bar). Alg. \ref{alg:adagrad} with regularization (purple) has similar performance as Alg. \ref{alg:adagrad} (red) and Alg. \ref{alg:adagrad} with Alg. \ref{alg:cleanup} (green) except for scenario 1 where there is a significant performance degradation compared to the other algorithms. We see also that performing the power reduction algorithm does not reduce the performance and might even lead to a small increase, as seen for scenario 3.
Fig. \ref{fig:power} shows the average transmit power of the different methods. We first see that the initialization has the smallest average power as it was initialized to the minimal transmit power of each device. The transmit power of Alg.~\ref{alg:adagrad} (red) is always the highest because there is no mechanism to try to limit the transmit power, in contrast with the green and purple bars. The average transmit power of Alg. \ref{alg:adagrad} with Alg. \ref{alg:cleanup} is always smaller than Alg. \ref{alg:cleanup} with $\ell^1$ regularization and is also smaller than the baselines. Thus the method using Alg. \ref{alg:adagrad} with Alg. \ref{alg:cleanup} yields a slot and power allocations that avoids undecodable collisions with a transmit power that is smaller than the baselines.  

\section{Conclusion}
To fully exploit the capability of a SIC-enabled base station, controlling the transmit power of the devices as well as with their transmit slot is important. In this paper, we proposed algorithms for jointly optimizing both slot and power allocations for all devices of the network. This increases the performance of the network, measured by the expected number of devices that transmit without outage. We presented a second method aimed at minimizing the transmit power of the devices without decreasing the throughput performances. Our simulations show that our algorithm can significantly outperform existing ALOHA-based algorithms in terms of both throughput and power consumption.

\bibliographystyle{IEEEtran}
\bibliography{SIC.bib, Aloha.bib, NetworkOptimization.bib, StochasticApproximation.bib, TechnologyReview.bib,torch.bib}

\begin{thebibliography}{10}
\providecommand{\url}[1]{#1}
\csname url@samestyle\endcsname
\providecommand{\newblock}{\relax}
\providecommand{\bibinfo}[2]{#2}
\providecommand{\BIBentrySTDinterwordspacing}{\spaceskip=0pt\relax}
\providecommand{\BIBentryALTinterwordstretchfactor}{4}
\providecommand{\BIBentryALTinterwordspacing}{\spaceskip=\fontdimen2\font plus
\BIBentryALTinterwordstretchfactor\fontdimen3\font minus
  \fontdimen4\font\relax}
\providecommand{\BIBforeignlanguage}[2]{{%
\expandafter\ifx\csname l@#1\endcsname\relax
\typeout{** WARNING: IEEEtran.bst: No hyphenation pattern has been}%
\typeout{** loaded for the language `#1'. Using the pattern for}%
\typeout{** the default language instead.}%
\else
\language=\csname l@#1\endcsname
\fi
#2}}
\providecommand{\BIBdecl}{\relax}
\BIBdecl

\bibitem{robertsALOHAPacketSystem1975}
L.~G. Roberts, ``{{ALOHA}} packet system with and without slots and capture,''
  \emph{ACM SIGCOMM Computer Communication Review}, vol.~5, no.~2, pp. 28--42,
  Apr. 1975.

\bibitem{wieselthierExactAnalysisPerformance1989b}
J.~Wieselthier, A.~Ephremides, and L.~Michaels, ``An exact analysis and
  performance evaluation of framed {{ALOHA}} with capture,'' \emph{IEEE
  Transactions on Communications}, vol.~37, no.~2, pp. 125--137, Feb. 1989.

\bibitem{paoliniCodedSlottedALOHA2015}
E.~Paolini, G.~Liva, and M.~Chiani, ``Coded {{Slotted ALOHA}}: {{A Graph-Based
  Method}} for {{Uncoordinated Multiple Access}},'' \emph{IEEE Transactions on
  Information Theory}, vol.~61, no.~12, pp. 6815--6832, Dec. 2015.

\bibitem{livaGraphBasedAnalysisOptimization2011}
G.~Liva, ``Graph-{{Based Analysis}} and {{Optimization}} of {{Contention
  Resolution Diversity Slotted ALOHA}},'' \emph{IEEE Transactions on
  Communications}, vol.~59, no.~2, pp. 477--487, Feb. 2011.

\bibitem{kalorRandomAccessSchemes2018a}
A.~E. Kalor, O.~A. Hanna, and P.~Popovski, ``Random {{Access Schemes}} in
  {{Wireless Systems}} with {{Correlated User Activity}},'' in \emph{2018
  {{IEEE}} 19th {{International Workshop}} on {{Signal Processing Advances}} in
  {{Wireless Communications}} ({{SPAWC}})}.\hskip 1em plus 0.5em minus
  0.4em\relax Kalamata: IEEE, Jun. 2018, pp. 1--5.

\bibitem{raghuwanshiChannelSchedulingIoT2024}
P.~Raghuwanshi, O.~L.~A. L{\'o}pez, P.~Popovski, and M.~{Latva-aho}, ``Channel
  {{Scheduling}} for {{IoT Access}} with {{Spatial Correlation}},'' \emph{IEEE
  Communications Letters}, pp. 1--1, 2024.

\bibitem{zhengStochasticResourceOptimization2021}
C.~Zheng, M.~Egan, L.~Clavier, A.~E. Kalor, and P.~Popovski, ``Stochastic
  {{Resource Optimization}} of {{Random Access}} for {{Transmitters With
  Correlated Activation}},'' \emph{IEEE Communications Letters}, vol.~25,
  no.~9, pp. 3055--3059, Sep. 2021.

\bibitem{zhengStochasticResourceAllocation2022}
------, ``Stochastic {{Resource Allocation}} for {{Outage Minimization}} in
  {{Random Access}} with {{Correlated Activation}},'' in \emph{2022 {{IEEE
  Wireless Communications}} and {{Networking Conference}} ({{WCNC}})}.\hskip
  1em plus 0.5em minus 0.4em\relax Austin, TX, USA: IEEE, Apr. 2022, pp.
  1635--1640.

\bibitem{jeannerotMitigatingUserIdentification2023}
A.~Jeannerot, M.~Egan, L.~Chetot, and J.-M. Gorce, ``Mitigating {{User
  Identification Errors}} in {{Resource Optimization}} for {{Grant-Free Random
  Access}},'' in \emph{2023 {{IEEE}} 97th {{Vehicular Technology Conference}}
  ({{VTC2023-Spring}})}, Jun. 2023, pp. 1--6.

\bibitem{jeannerotExploitingDeviceHeterogeneity2024}
A.~Jeannerot, M.~Egan, and J.-M. Gorce, ``Exploiting {{Device Heterogeneity}}
  in {{Grant-Free Random Access}}: {{A Data-Driven Approach}},'' \emph{IEEE
  Transactions on Vehicular Technology}, pp. 1--11, 2024.

\bibitem{chetotJointIdentificationChannel2021}
L.~Chetot, M.~Egan, and J.-M. Gorce, ``Joint {{Identification}} and {{Channel
  Estimation}} for {{Fault Detection}} in {{Industrial IoT With Correlated
  Sensors}},'' \emph{IEEE Access}, vol.~9, pp. 116\,692--116\,701, 2021.

\bibitem{mounirSelectionPowerAllocation2022}
M.~Mounir, M.~B. El\_Mashade, and A.~Mohamed~Aboshosha, ``On {{The Selection}}
  of {{Power Allocation Strategy}} in {{Power Domain Non-Orthogonal Multiple
  Access}} ({{PD-NOMA}}) for {{6G}} and {{Beyond}},'' \emph{Transactions on
  Emerging Telecommunications Technologies}, vol.~33, no.~6, p. e4289, 2022.

\bibitem{hoglundOverview3GPPRelease2017}
A.~Hoglund, X.~Lin, O.~Liberg, A.~Behravan, E.~A. Yavuz, M.~Van Der~Zee,
  Y.~Sui, T.~Tirronen, A.~Ratilainen, and D.~Eriksson, ``Overview of {{3GPP
  Release}} 14 {{Enhanced NB-IoT}},'' \emph{IEEE Network}, vol.~31, no.~6, pp.
  16--22, Nov. 2017.

\bibitem{duchiAdaptiveSubgradientMethods2011}
J.~Duchi, E.~Hazan, and Y.~Singer, ``Adaptive {{Subgradient Methods}} for
  {{Online Learning}} and {{Stochastic Optimization}},'' \emph{Journal of
  Machine Learning Research}, p.~39, 2011.

\bibitem{Paszke_PyTorch_An_Imperative_2019}
A.~Paszke, S.~Gross, F.~Massa, A.~Lerer, J.~Bradbury, G.~Chanan, T.~Killeen,
  Z.~Lin, N.~Gimelshein, L.~Antiga, A.~Desmaison, A.~Kopf, E.~Yang, Z.~DeVito,
  M.~Raison, A.~Tejani, S.~Chilamkurthy, B.~Steiner, L.~Fang, J.~Bai, and
  S.~Chintala, ``{PyTorch: An Imperative Style, High-Performance Deep Learning
  Library},'' in \emph{Advances in Neural Information Processing Systems 32},
  H.~Wallach, H.~Larochelle, A.~Beygelzimer, F.~d'Alché Buc, E.~Fox, and
  R.~Garnett, Eds.\hskip 1em plus 0.5em minus 0.4em\relax Curran Associates,
  Inc., 2019, pp. 8024--8035.

\end{thebibliography}
\end{document}